\documentclass[aps,prl,twocolumn,groupedaddress]{revtex4-1}
\usepackage{graphics}
\usepackage{graphicx}
\usepackage{algorithm}
\usepackage{algcompatible}
\usepackage{float}
\begin{document}

%\preprint{}

\title{A novel approach to the study of critical systems}


\author{Lorenzo \surname{Palmieri}}
\email{l.palmieri16@imperial.ac.uk}
\affiliation{Centre for Complexity Science and Department of Mathematics, Imperial College London, South Kensington Campus, SW7 2AZ, UK;}

\author{Henrik Jeldtoft \surname{Jensen}}
\email{h.jensen@imperial.ac.uk}
\affiliation{Centre for Complexity Science and Department of Mathematics, Imperial College London, South Kensington Campus, SW7 2AZ, UK;}
\affiliation{Institute of Innovative Research, Tokyo Institute of Technology, 4259, Nagatsuta-cho, Yokohama 226-8502, Japan Japan.}

%\date{\today}

\begin{abstract}
We introduce a novel approach to study the critical behavior of equilibrium and non-equilibrium systems which is based on the concept of an instantaneous correlation length. We analyze in detail two classical statistical mechanical systems: the XY model and the Ising model, and one of the prototype models of Self-Organized Criticality: the forest fire model (FFM). The proposed method can both capture the critical behavior of the XY model and the Ising model and discriminate between the nature of the phase transition in the two scenarios. When applied to the FFM, it gives surprising results, suggesting that the model could be critical despite displaying broken scaling in the distribution of cluster sizes. 
\end{abstract}

%\keywords{}

\maketitle
\section{Introduction}
The concept of criticality is widely used in many disciplines, spanning finance\citep{Crashes2000, Biondo2015, Sornette2015, SornetteWSMC}, meteorology \cite{Peters2006, Yano2012, Selvam2017}, neuroscience\cite{Cocchi2017, Hesse2014, Iyer2018, Brochini2016} and physics \cite{Sornette2006, Markovic2014, GunnarBook, Moloney2005}. A system in a critical state is usually characterized by scale invariance and self-similarity, and by the development of strong instabilities which are caused by the emergence of long-range temporal or spatial interactions. In statistical physics, the term criticality indicates the behavior of a system near a critical point, which is typically associated with a phase transition between two different states. A classic example of phase transition is the behavior of magnets near a critical temperature $T_c$, which separates an ordered state at low temperatures ($T<T_c$) from a disordered one at high temperatures ($T>T_c$). When a system is in a critical state, it is highly susceptible to external perturbations, and it is characterized by the emergence of long-range correlations between its constituent components. This high susceptibility is a direct consequence of the self-similarity of the correlation function, which emerges from microscopic interactions and leads to the presence of strong correlations on all scales of the system. 
\\ Formally, correlations are described by the covariance between two microscopic physical quantities. In statistical physics, the correlation function is usually defined as the difference between the canonical ensemble average $\langle ... \rangle$ of the scalar product between two random variables $s_1$ and $s_2$ (usually spins or particles) at positions $r_0$ and $r_0 + r$ and their uncorrelated average product:
\begin{equation} \label{termo}
C(r) = \langle s_1(r_0) s_2(r_0 + r) \rangle - \langle s_1(r_0) \rangle \langle s_2(r_0 + r) \rangle
\end{equation}
Introducing the external control parameter $X$, at a critical point $X_c$ and in the thermodynamic limit one expects to find scale-invariance in the correlations, which corresponds to a power-law behavior of the correlation function:
\begin{equation}\label{critC}
C(r|X_c) \sim r^{-\eta}
\end{equation}
Eq. \ref{critC} implies that correlations behave in the same way for any arbitrary rescaling of the distance by a factor $\mu$, i.e. if $r \rightarrow \mu r$ then one still has $C(r|X_c) \sim (\mu r)^{-\eta} \sim r^{-\eta}$.  The fact that correlations are present at all scales translates in long range correlations and the resulting critical behavior of the whole system. Away from the critical point, correlations typically decay as an exponential function, and the characteristic length of the exponential is referred to as the correlation length $\bar{\xi}$. The typical functional form that is assumed for the correlation function near a critical point is
\begin{equation}
C(r|X) \sim r^{-\eta}r^{-\frac{r}{\bar{\xi}(X,L)}}
\end{equation}
where $\bar{\xi}$ indicates the typical length over which two agents are correlated and depends on the control parameter of the system $X$ and the system size $L$. This length is limited by $L$ and diverges in the thermodynamic limit in correspondence of the critical value of the control parameter $X_c$, i.e. $\bar{\xi}(X_c,L\rightarrow \infty) \rightarrow \infty$, giving Eq.\ref{critC}. It is clear then that the correlation length act as a parameter that describes the typical extension of correlations inside a system and therefore represents the most reasonable quantity to look at when one investigates the critical behavior of a physical system. However, it is essential to observe that measuring a diverging correlation length is not enough to determine if a system is in a critical state, because it does not convey any information about the scaling behavior of the system. In other words, one could observe a divergent correlation length even in a system that is not scale-invariant and therefore not critical, as will be discussed in the next sections. 
\\ In this paper, we introduce a new method to investigate the critical behavior of a system. This method is still based on the study of the correlation function, but introduces a new correlation length that is no longer a parameter of the system, but a stochastic variable which distribution is able to catch at the same time the scale-invariance of the system, the asymptotic behavior of the correlation length and the universal properties of the model.

\section{The instantaneous correlation length formalism}
The method proposed is based on the instantaneous correlation length introduced in \cite{WC}.  For simplicity, we consider models defined on a 2D lattice from which we sample $N$ independent lattice configurations $S_1, S_2,  \ldots S_N$ during the time evolution. The classic estimate of the correlation length $\bar{\xi}$ goes as follows: for each configuration $S_t$ one computes the two-point correlation function $C_t (r)$ between two spins $s_1$ and $s_2$ at positions $r_0$ and $r_o + r$. Assuming translational invariance, one has:
\begin{equation}
C_t(r) = \langle s_1(r_0) s_2(r_0 + r) \rangle_t - \langle s(r_0) \rangle_t ^2
\end{equation}
where the average $\langle \ldots \rangle$ is taken summing over all the possible pairs of spins and values of $r_0$ at a time $t$. Iterating this procedure for different configurations, one obtains an ensemble of correlation functions $\lbrace C_1, C_2, \ldots, C_N \rbrace$, which can be used to compute the time-averaged correlation function $\overline{C}(r)$ for which the following functional form is usually assumed near a critical point:
\begin{equation} \label{fit}
\overline{C}(r) \sim r^{-\eta} e^{-\frac{r}{\bar{\xi}(X)}} 
\end{equation}
where $\bar{\xi}(X)$ is the correlation length which depends on a control parameter $X$. In correspondence of the critical value of the control parameter $X=X_c$, the correlation length $\bar{\xi}(X_c)$ diverges in the thermodynamic limit, and the correlation function decays algebraically. 
\\Now we introduce the instantaneous correlation length formalism. Assuming that the system size is sufficiently large to give reasonable statistics for the instantaneous correlation function $C_t(r)$, one can fit the instantaneous correlation length $\xi_t$ using the same functional form that is used in Eq. \ref{fit}. Doing this, one obtains an ensemble of instantaneous correlation lengths $\lbrace \xi_1, \xi_2, \ldots, \xi_N \rbrace$. Each $\xi_t$ is a measure of how $critical$ a single configuration $S_t$ is. If the system is far from a critical point, one expects $\xi_t$ to be always small because the correlation function will decrease exponentially fast. On the other hand, as the system approaches $X_c$ there will be an increasing fraction of configurations with a big correlation length, which corresponds to a power-law behavior of the correlation function. Although one expects the ensemble averaged correlation length $\bar{\xi}(X)$ and the average instantaneous correlation length $\xi$ to scale in the same way, it is essential to stress the fact that they are two distinct mathematical objects: the first being a parameter of the ensemble averaged correlation function and the second being a stochastic variable. Indeed, the strength of this new approach lies in the fact that we can now use the ensemble of $\xi_t$ to compute not only the average correlation length $\langle \xi \rangle$, but also the distribution of correlation lengths $P(\xi)$. $P(\xi)$ is an entirely new physical object and, as we will see, contains plenty of information about the critical behavior of the system under analysis. 
\section{The distribution of the instantaneous correlation lengths}
Using $P(\xi)$, it is possible to determine whether a system is at a critical point or not. This can be done by looking at the conditional probability $P(\xi|X)$, which should become scale invariant in correspondence of $X_c$. Assuming simple scaling, one expects:
\begin{equation} \label{Scansatz}
P(\xi, \xi_c) \propto G \Big(\frac{\xi}{\xi_c} \Big) \xi^{-\tau}
\end{equation}
for $\xi$ and $\xi_c$ bigger than a constant lower cut-off $\xi_0$.  In Eq.\ref{Scansatz}, $\xi_c$ represents an upper cut-off that diverges in the thermodynamic limit, $ G \Big(\frac{\xi}{\xi_c} \Big) $  is a universal scaling function and $\tau$ is a critical scaling exponent. In general, the upper cut-off scales as $\xi_c \sim aL^\beta$, where $a$ is a non-universal metric factor and $\beta$ is related to the universal spatial dimension of the observable \citep{Domb,Christensen2008}. From Eq. \ref{Scansatz} one can compute the $n^{th}$ moment as 
\begin{equation} \label{Moments}
\langle \xi^n \rangle = \xi_c^{n-\tau+1}\int_{\frac{\xi_0}{\xi_c}}^{\infty} G \big( u \big) u^{n-\tau} du
\end{equation}
Imposing normalization ($n=0$) one gets $\tau \geq 1$. If one absorbs the non-universal constant $a$ in the definition of $G(u)$ and assumes that integral in Eq. \ref{Moments} converges in zero, then in the thermodynamic limit the average correlation length is given by
\begin{equation} \label{xilim}
\langle \xi \rangle = L^{\beta(2-\tau)}\int_{0}^{\infty} G \big( u \big) u^{1-\tau} du
\end{equation}
In the following sections, we will study the behavior of $P(\xi)$ in two traditional statistical mechanical systems, the Ising Model and the XY Model, and to one of the prototype models of Self-Organized Criticality, the forest Fire Model. We conclude this section observing that if Eq. \ref{Scansatz} holds, then it automatically allows the introduction of the new critical exponent $\tau$.
\section{Ising Model}
\begin{figure}[h]
\centering
\includegraphics[width=9cm,height=9cm]{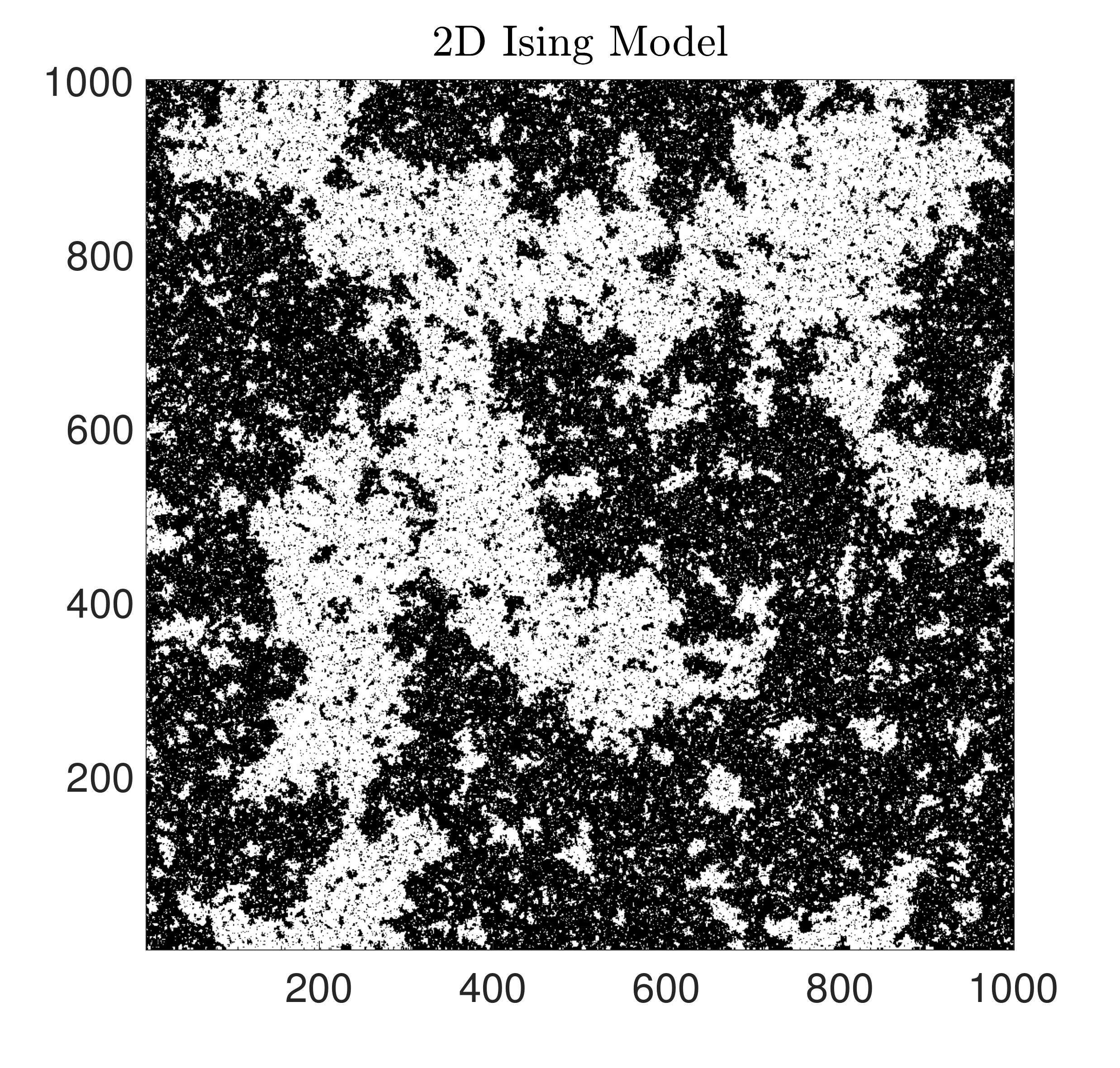}
\caption{Sample configuration of the 2D Ising model for a square lattice with $L=1000$ at $T_c$. }
\label{Isingspin}
\end{figure}

\begin{figure}[h]
\centering
\includegraphics[width=9cm,height=9cm]{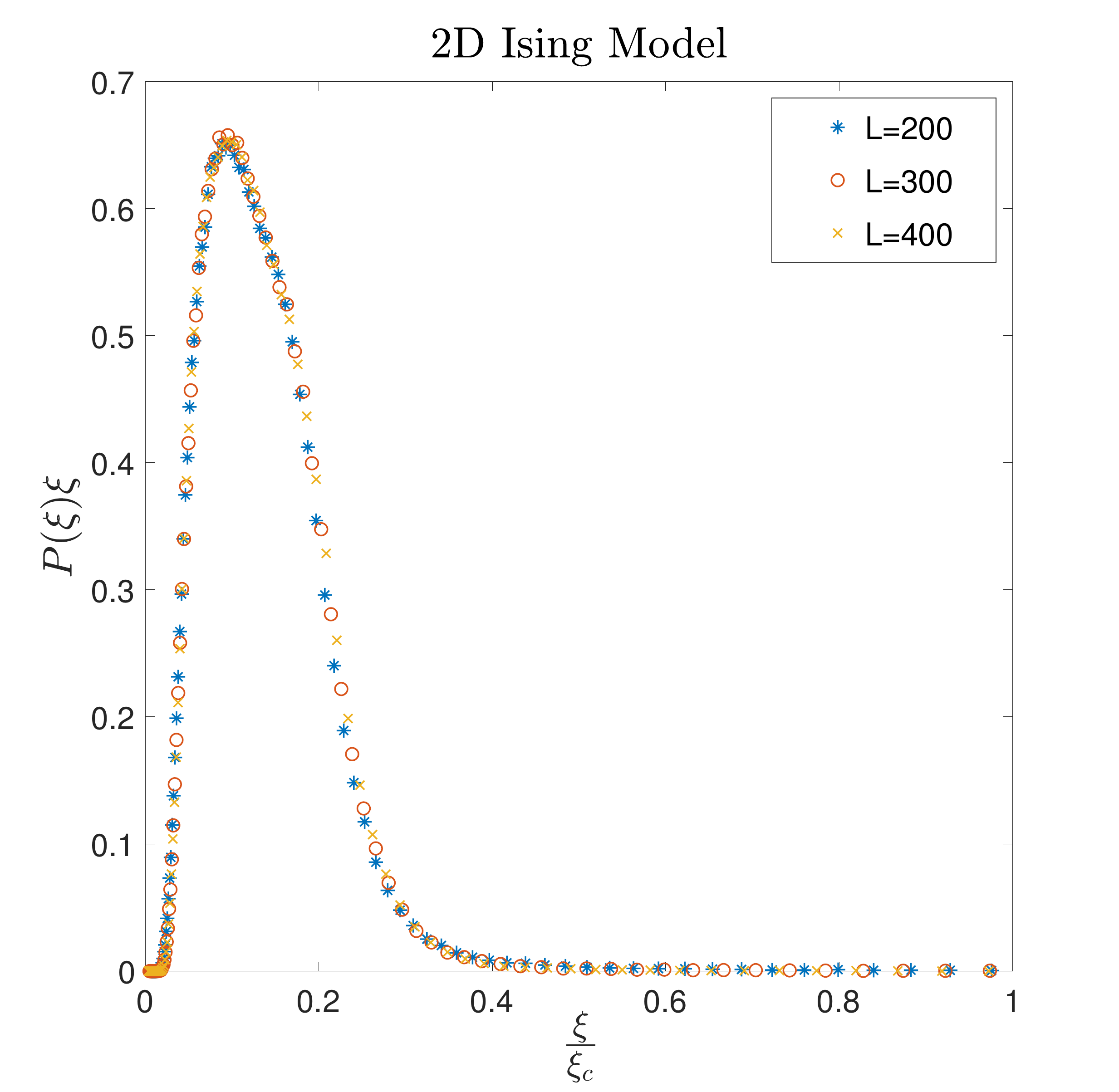}
\caption{Universal function $G(u)$ for the Ising model. A data collapse is only possible in correspondence of $T_c$ and for $\tau=1$ and $\beta=1$. }
\label{Ising}
\end{figure}

\begin{figure}[h]
\centering
\includegraphics[width=9cm,height=9cm]{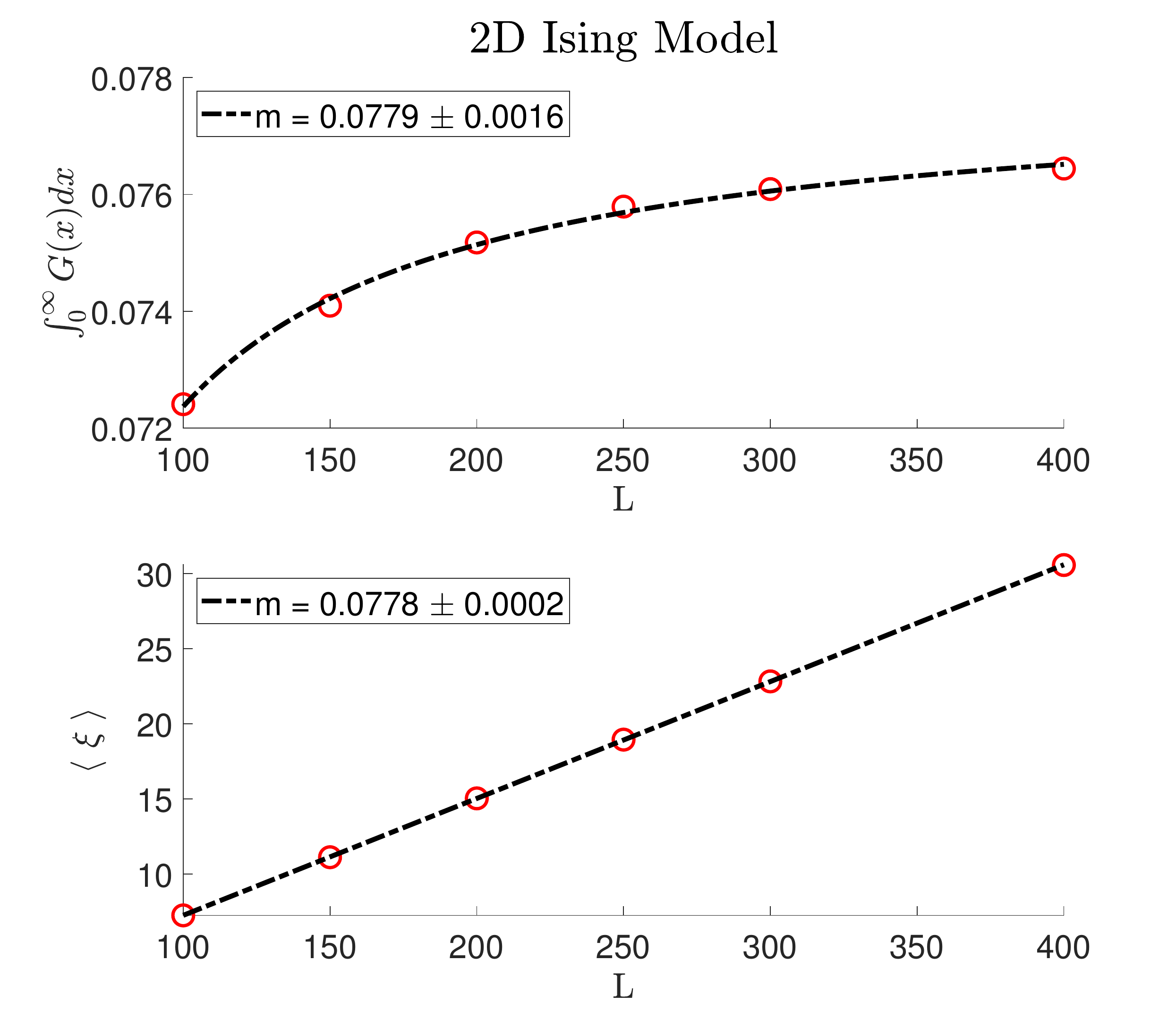}
\caption{Estimate of the constant of proportionality $m$ between $\langle \xi \rangle$ and $L$. In the upper panel,  $m$ has been estimated by fitting with a power-law the integral of the universal function $G(x)$ as a function of $L$. In the lower panel, $m$ has been estimated by means of a linear fit of $\langle \xi \rangle$ vs $L$. The integral in the upper panel converges towards the asymptotic value as $L^{-\lambda}$, with $\lambda = 1\pm 0.5$. The error corresponds to confidence bounds of $95\%$.}
\label{Isingfit}
\end{figure}

The Ising model is a mathematical model of ferromagnetism that was invented by Wilhelm Lenz in 1920 and solved for the first time in one dimension by Ernst Ising in 1925 \cite{1925Ising, HistoryIsing}. The model consists of $N$ interacting two-state spin variables $\sigma_i \in [-1,1]$ which represent adjacent magnetic dipoles. The energy that is associated with a given macro-configuration $\sigma$ is given by
\begin{equation}
H(\sigma) = -\sum_{<i,j>} J_{ij}\sigma_i \sigma _j - \sum_{i=1}^N h_i \sigma_i
\end{equation}
where the first sum is over all pairs of adjacent spins $<i,j>$, $J_{ij}$ is the interaction strength, and $h_i$ is the external magnetic field. The two-dimensional square lattice Ising model was solved in 1944 by Onsager \cite{Onsager} in the case of no external field  ($h_{i}=0$)  and assuming periodic boundary conditions and constant interaction strength along the x-axis ($J_x$) and the y-axis ($J_y$).  The 2D Ising Model is central in statistical physics because it is one of the simplest statistical models to exhibit a phase transition between an ordered phase (low temperatures) and a disordered phase (high temperatures). In the case of isotropic interactions $J_x=J_y$, the critical value of the temperature $T=T_c$ that marks the phase transition is given by 
\begin{equation} \label{onsager}
\frac{k_B T_{c}}{J}=\frac{2}{\ln(1+\sqrt{2})} \simeq 2.269
\end{equation}
If one looks at the lattice at different temperatures, it can be noted that the high-temperature phase is characterized by disorder, because the entropy introduced in the system by the temperature destroys long-range correlations, which results in random configurations with roughly half of the spins up and half of the spins down and no emergent complex structures. On the other hand, at low temperatures, most of the spins will be able to align in order to minimize the energy, giving rise to ordered configurations. At the critical point, the correlation function decays as a power-law, and we are in the presence of long-range interactions and the creation of fractal structures, as can be seen in Fig.\ref{Isingspin}. This behavior is reflected by the fact that the correlation length becomes proportional to the system size $L$ and diverges in the limit $L \rightarrow \infty$. 
In terms of the instantaneous correlation length formalism, if we assume that the upper cut-off scales like $L$, i.e. $\beta=1$, then in order to maintain the linear relationship between the system size $L$ and $\xi$, we should have $\tau=1$ in Eq. \ref{xilim},  meaning that the constant of proportionality would be given by the integral of the universal function $G(u)$. Under these assumptions and in the large $L$ limit, Eq. \ref{xilim} reduces to 
\begin{equation} \label{bella}
\langle \xi \rangle = L\int_{0}^{\infty} G \big( u \big) du
\end{equation}
As $\tau=1$ implies that the limit $\lim\limits_{u\to0} G(u) =0$ \cite{Christensen2008}. In order to verify if the theory is correct, we need to evaluate whether there is a value of the control parameter $X=X_c$ such that $P(\xi|X_c)$ becomes scale invariant. For the $2D $ Ising model, this corresponds to the critical temperature given by Onsager's solution in Eq. \ref{onsager}. In our simulations, we used the Wolff Algorithm \cite{Wolff1989} in order to reduce the critical slowing down and for each system size $L$ we sampled $10^6$ independent configurations to estimate $P(\xi)$ at $T_c$. In the following, we will use $\xi_0 = 1$ and $\xi_c=\frac{L}{\sqrt{2}}$ as the reference upper cut-off, as this is the maximum physical distance between two points on a square lattice with periodic boundary conditions. Plotting $P(\xi) \xi$ as a function of $\frac{\xi}{\xi_c}$, we can perform a data collapse in correspondence of $T_c$ (Fig.\ref{Ising}). The resulting curve corresponds to the universal scaling function $G(u)$ which according to Eq. \ref{xilim}, can be used to compute the proportionality constant between $\xi$ and $L$.  In Fig. \ref{Isingfit} it is shown how the integral of the universal function $G(x)$ converges to a value that is consistent with the estimated gradient of the line $\langle \xi \rangle = m L$. 
\\ In summary, when applied to the 2D Ising model, our method was able to identify the critical temperature as the $T$ for which $P(\xi|T)$ becomes scale invariant and to capture the scaling behavior of $\langle \xi \rangle$, which is consistent with the classical theory. In addition to these two well-known results, we were able to introduce a new critical exponent for the Ising model, i.e. $\tau=1$, and to relate the rate of growth of the correlation length to the universal function $G(u)$.  

\section{XY Model}
\begin{figure}[h]
	\centering
	\includegraphics[width=9cm,height=9cm]{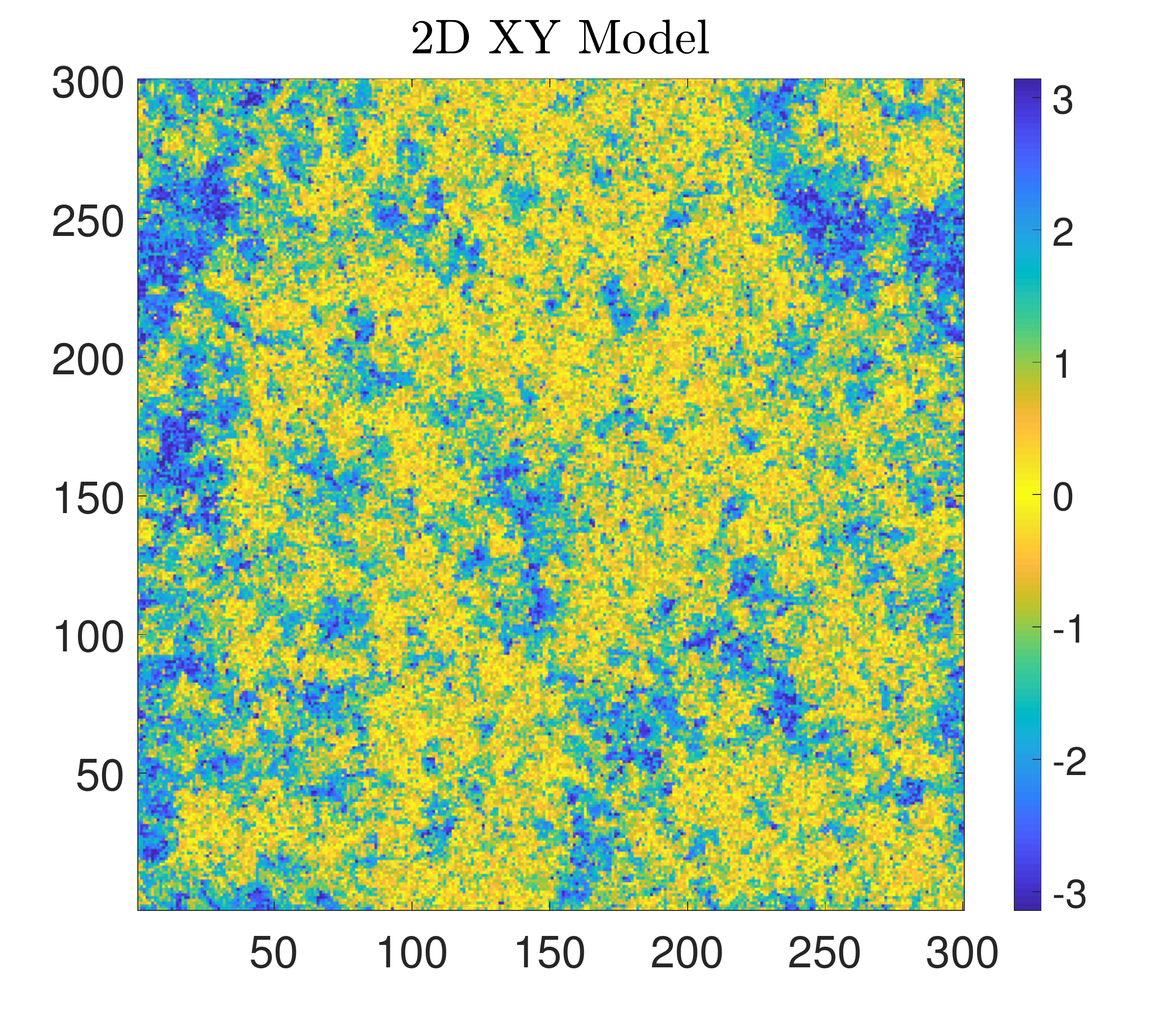}
	\caption{Sample configuration of the 2D XY model with L=300 at $T_{KT}$.}
	\label{XYsample}
\end{figure}

\begin{figure}[H]
\centering
\includegraphics[width=9cm,height=9cm]{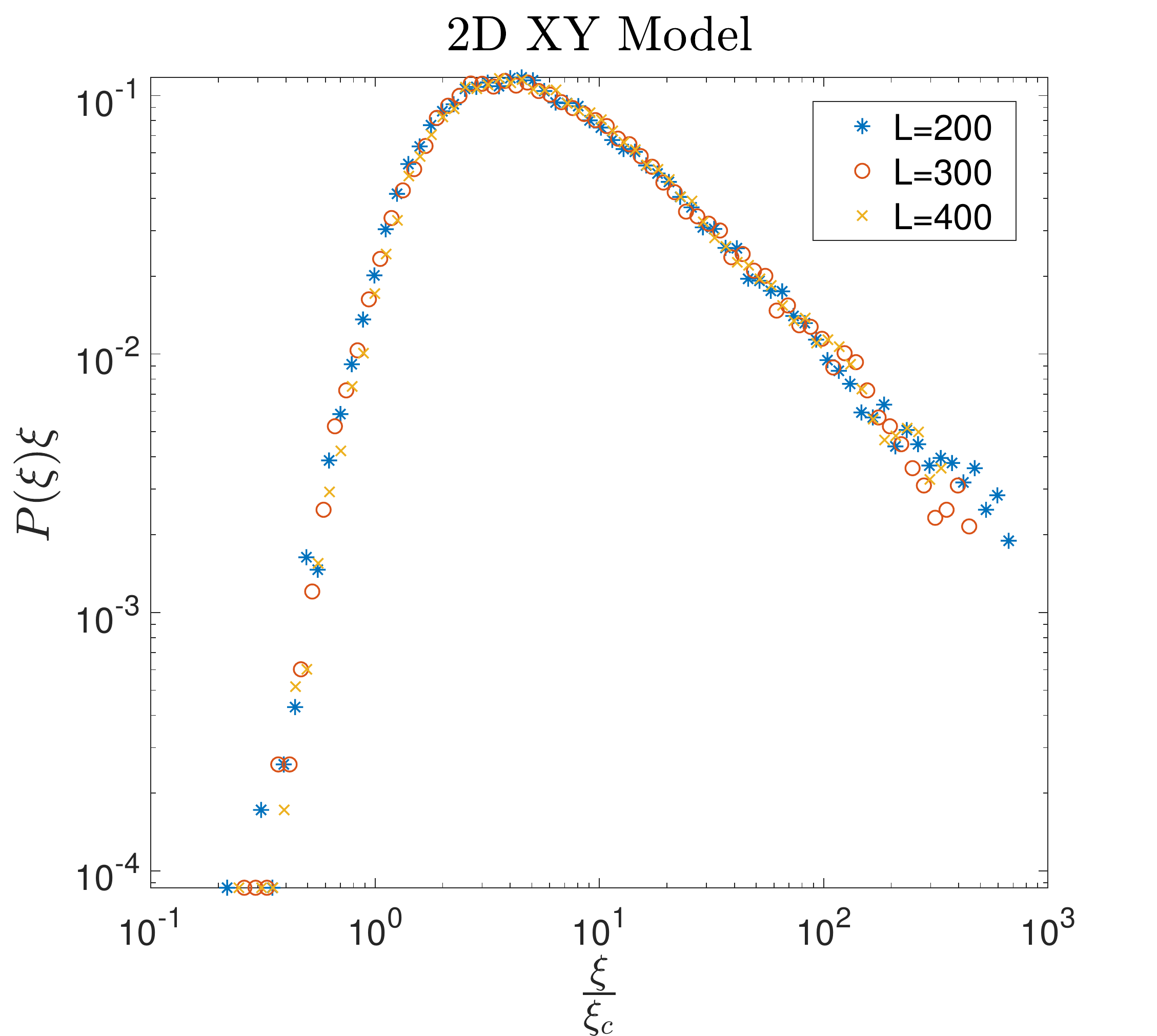}
\caption{Universal function $G(u)$ for the XY Model. The distribution of correlation lengths $P(\xi)$ becomes scale invariant in correspondence of $T_{KT}$ and for $\tau=1$ and $\beta=1$.}
\label{XY}
\end{figure}

The two-dimensional XY-model is a paricular case of the Heisenberg model, which was introduced in 1928 \cite{Heisenberg1928} as a model for ferromagnetism. Similarly to the Ising Model, it consists of a system of spins in a lattice with the difference that the individual spins can rotate in any direction and are not constrained to take only two values. The energy of the model is given by
\begin{equation}
H(\sigma) = -\sum_{<i,j>} J_{ij} \vec{\sigma_i} \cdot \vec{\sigma _j} = -\sum_{<i,j>} J_{ij} cos(\theta_i - \theta_j)
\end{equation}
where the first sum is over pairs of adjacent spins $<i,j>$, $J_{ij}$ is the interaction strength, and $\theta_i$ is the angle that a spin $\vec{\sigma_i}$ makes with respect to some arbitrary direction in the lattice plane. As for the Ising model, in our simulation we keep the interaction strengths constant $J_{ij}=J$ and apply periodic boundary conditions. A typical realization of the model is represented in Fig. \ref{XYsample}. 
The two-dimensional version of this model is of particular interest because at high temperatures correlations decay exponentially fast, while at low temperatures they decay with a power-law, even though in both cases the overall magnetization is zero. This peculiar transition is named after Kosterlitz and Thouless who first discovered it in 1973 \cite{Kosterlitz_1973}. The XY-model is a relevant case to discuss in this context because of the behaviour of the correlation length, which diverges even for finite systems at temperatures below the Kosterlitz-Thouless temperature $T_{KT} \simeq 0.892 J$ \cite{Olsson1995, Janke1993,Hasenbusch2005}. 
In the XY model, the two-point correlation function is defined as \citep{Kosterlitz1974}
\begin{equation}
C(r) = \langle\cos(\theta(r_0)-\theta(r_0 + r) )\rangle
\end{equation}
In our simulations we used the Wolff algorithm \citep{Wolff1989} sampling $10^5$ independent configurations to estimate $P(\xi)$ at $T_{KT}$ and used $\xi_0=1$ and $\xi_c=\frac{L}{\sqrt{2}}$. As for the Ising model, it is possible to perform a data collapse for $P(\xi)$  in correspondence of $T_{KT}$ and for $\tau=1$ and $\nu=1$. Although the Ising model and the XY model share the same exponents, we can observe in Fig. \ref{XY} that in the XY model, $\xi$ is able to exceed the system size $L$. This is in line with the theory, which predicts a pure power-law in two dimensions in correspondence of $T_{KT}$ \cite{chaikin1995}. The presence of the Kosterlitz-Thouless phase transition and the behavior of the correlations is summarized in Fig. \ref{XY2}, where we plot the conditional probability $P(\xi>L)$ at different temperatures. As one lowers the temperature, the fraction of correlation lengths that exceed the system size goes from $0$ to $90 \%$, which corresponds to the pure power-law decay of correlations at $T<T_{KT}$.
%La P(xi) e' scale invariant. Eta invece e' tipo Gaussiana e si restringe intorno a un valor medio di circa 0.18. non e' universale. 

\begin{figure}[H]
\centering
\includegraphics[width=9cm,height=9cm]{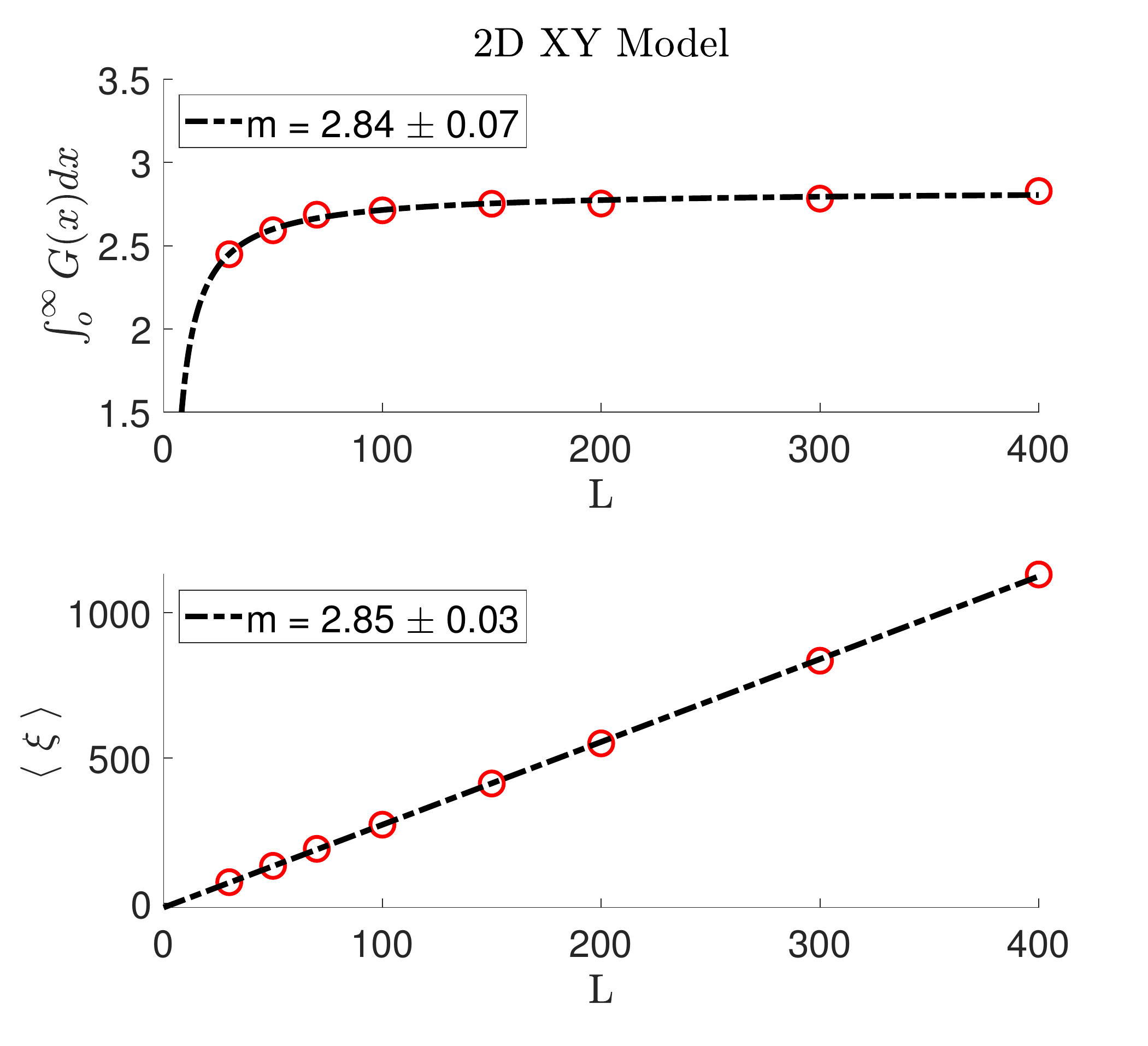}
\caption{Estimate of the constant of proportionality $m$ between $\langle \xi \rangle$ and $L$. In the upper panel, $m$ is estimated fitting the integral of the universal function $G(x)$ with a power-law, while in the lower panel it is fitted directly from $\langle \xi \rangle$. The integral in the upper panel converges towards the asymptotic value as $L^{-\lambda}$, with $\lambda = 1\pm 0.5$. The error corresponds to confidence bounds of $95\%$.}
\label{XY_ave}
\end{figure}

\begin{figure}[H]
\centering
\includegraphics[width=9cm,height=9cm]{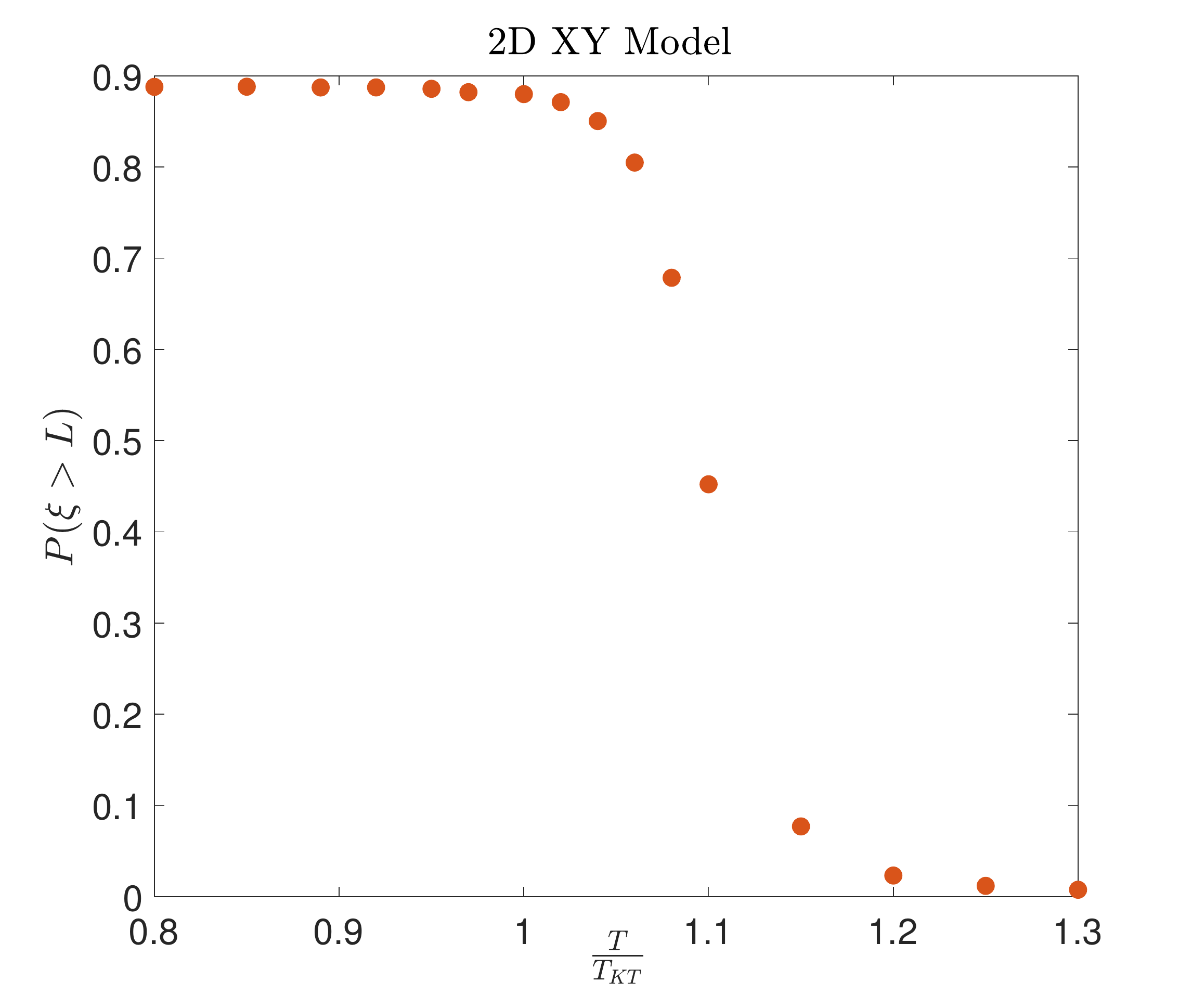}
\caption{Kosterlitz-Thouless phase transition in the 2D XY model with $L=100$. The fraction of correlation lengths that exceed the system size goes from $0$ to $1$ as the temperature approaches $T_{KT}$.}
\label{XY2}
\end{figure}

\section{Forest Fire Model}
\begin{figure}[h]
\centering
\includegraphics[width=9cm,height=9cm]{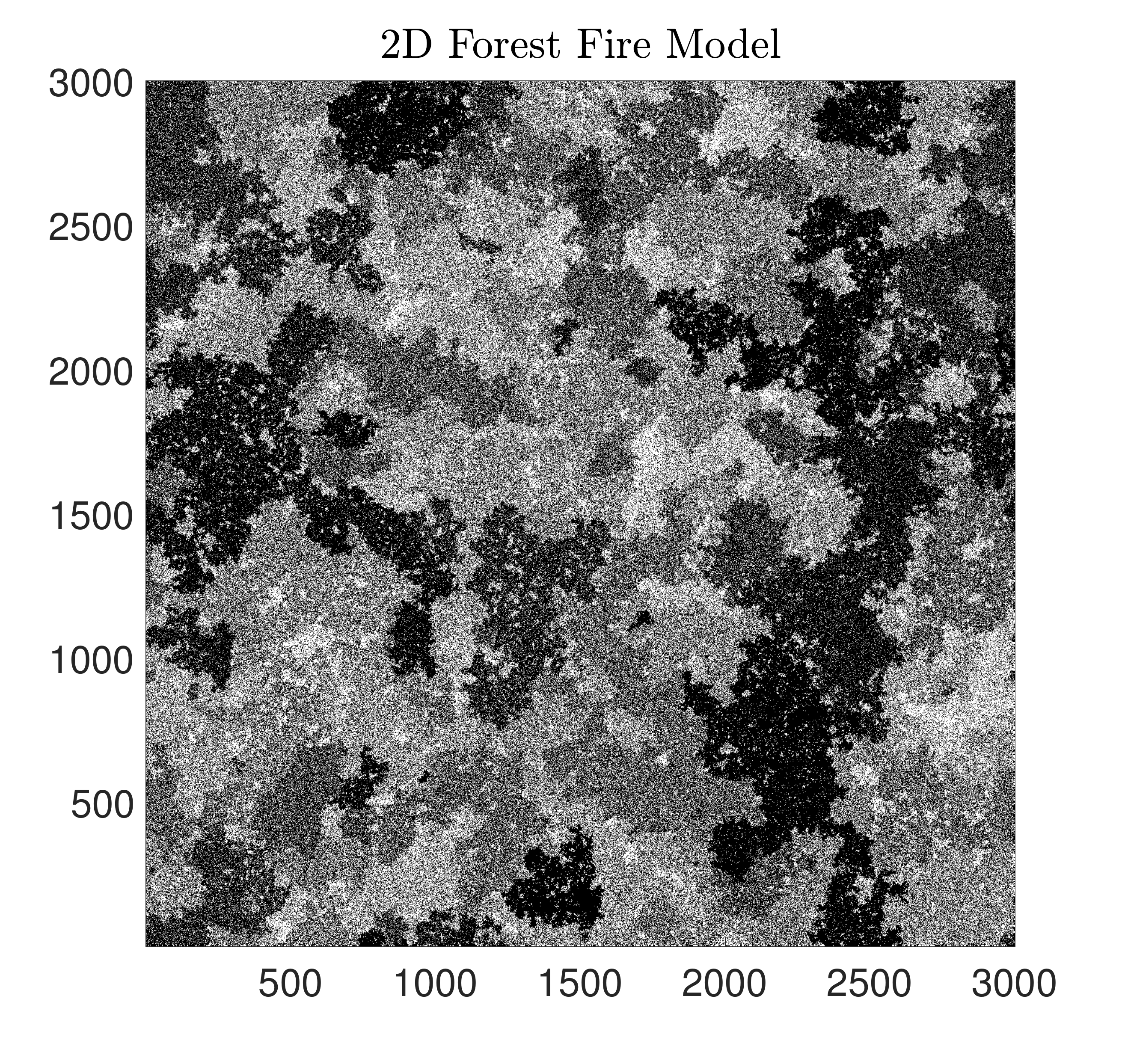}
\caption{Sample configuration of the Forest Fire Model for L=3000.}
\label{FFM_sample}
\end{figure}

The last model we consider is one of the prototype models of Self Organized Criticality: the Drossel$-$Schwabl Forest Fire Model (FFM) \cite{DS-FFM}. This model is different form the Ising Model and the XY model because it entails a dissipative dynamics and does not have an external control parameter, like temperature, that can be fine-tuned in order to reach a critical state. The dynamic involves the occupation of empty sites on a 2D grid with new trees (planting steps) and the removing of entire clusters of trees (burning steps). The creation of new trees and the removal of clusters results in the typical patchy appearance of the lattice, which is characterized by the presence of patches of different densities (Fig. \ref{FFM_sample}).
The way we implement the FFM follows \cite{Grassberger1993, Clar1994, Schenk2000, Gunnar2002} and is concisely summarized by the following pseudo-code:
\begin{algorithm}[H]
	\caption{Forest Fire Model}
	\label{Algo}
	\begin{algorithmic} [0]
		\WHILE{True}
		\FOR{i=1:$\theta$} \STATE{choose randomly a site s} 
		\IF{s is empty} \STATE {s becomes occupied}  \ENDIF 
		\ENDFOR
		\STATE{choose randomly a site s}
		\IF{s is occupied}
		\STATE {collect statistics}
		\STATE {burn the whole cluster connected to s}
		\ENDIF 
		\ENDWHILE
	\end{algorithmic}
\end{algorithm}
To estimate $P(\xi)$, we collected $10^6$ independent configurations after a transient of $5\cdot10^6$ burning steps. From Alg. \ref{Algo} it is clear that two parameters must be considered: the number of trees that one tries to plant $\theta$, and the system size $L$. In order to reach a critical state one would like to have both $L$ and $\theta$ infinitely large, although there is not a clear rule about how to tune $\theta$ for a finite system, and in the literature different authors have used quite a large span of $\theta$ values for the same systems size $L$ \cite{Gunnar2002,Grassberger2002}. Despite the model being introduced as critical, it was subsequently realized that the observed power-law in the distribution of clusters sizes displayed deviations from perfect scaling for large system sizes \cite{Grassberger2002, Gunnar2002}, implying that the model is not critical in the sense of being scale-free \cite{Gunnar2002}, and that all proposed scaling laws seem to be just transient \citep{Grassberger1993}. The correlation length was first studied in \cite{Henley1993} for systems sizes $L$ and $\theta$ up to $L=512$ and $\theta= 2048$, finding that $\xi \sim \theta^\nu$, with $\nu=0.56$. The authors also studied the connected correlation function finding $\nu_c = 0.58$, and attributed this discrepancy between the two exponents to numerical error. Another estimate for larger system sizes was given in \cite{Honecker1997}, where the authors used up to $L=17408$ and $\theta=10^4$ finding $\nu=0.541$ and $\nu_c=0.576$ to be statistically inconsistent, and therefore concluding that the model presents two different diverging correlation lengths. This finding points in the same direction as the lack of scaling observed in the distribution of cluster sizes. However, as it was noted in \citep{Grassberger1993}, there seem to be small deviations from a power-law in Fig.1 of \cite{Honecker1997}, meaning that the estimate of $\nu$ would be unreliable and therefore not suitable to confirm the presence of multiple diverging correlation lengths. 
 \\ Now we want to apply the instantaneous correlation length formalism to investigate whether $P(\xi)$ displays broken scaling as one should expect from a non-critical model. A similar approach was adopted in \citep{WC}, where the critical exponent was obtained by fitting the tail of $P(\xi)$. However, the tail includes contributions from the universal function $G(\xi,L)$ and therefore that estimate of the critical exponent is spurious. 
\subsection{Critical Behavior in the Forest fire Model}
\begin{figure}[h]
\centering
\includegraphics[width=9cm,height=9cm]{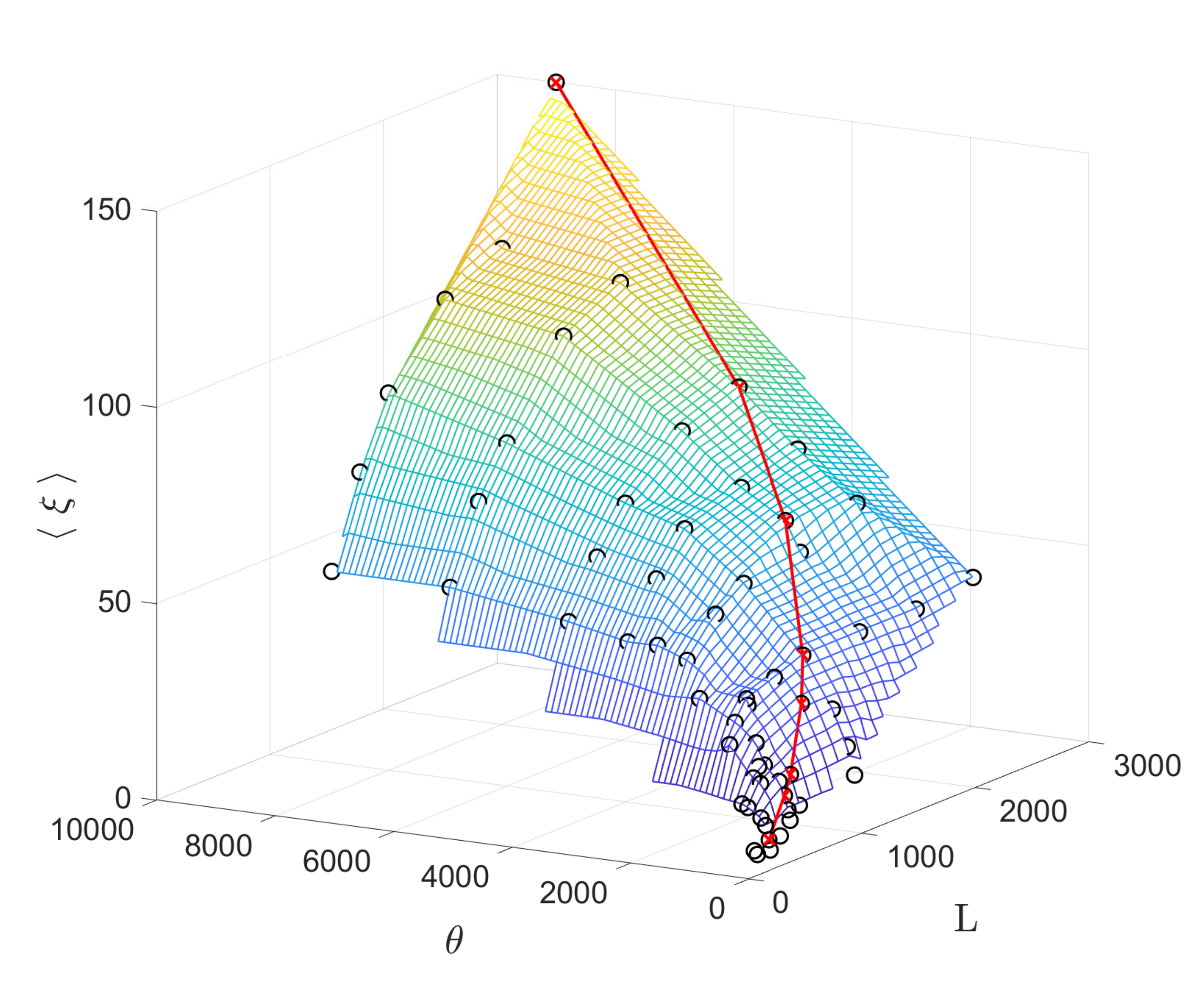}
\caption{$\langle \xi \rangle$ as a function of $\theta$ and $L$. The black dots correspond to the numerical simulations that have been performed to extrapolate the surface, while the red line corresponds to the path $\frac{\theta}{L^2}=10^{-3}$.}
\label{surf}
\end{figure}

\begin{figure}[h]
\centering
\includegraphics[width=9cm,height=9cm]{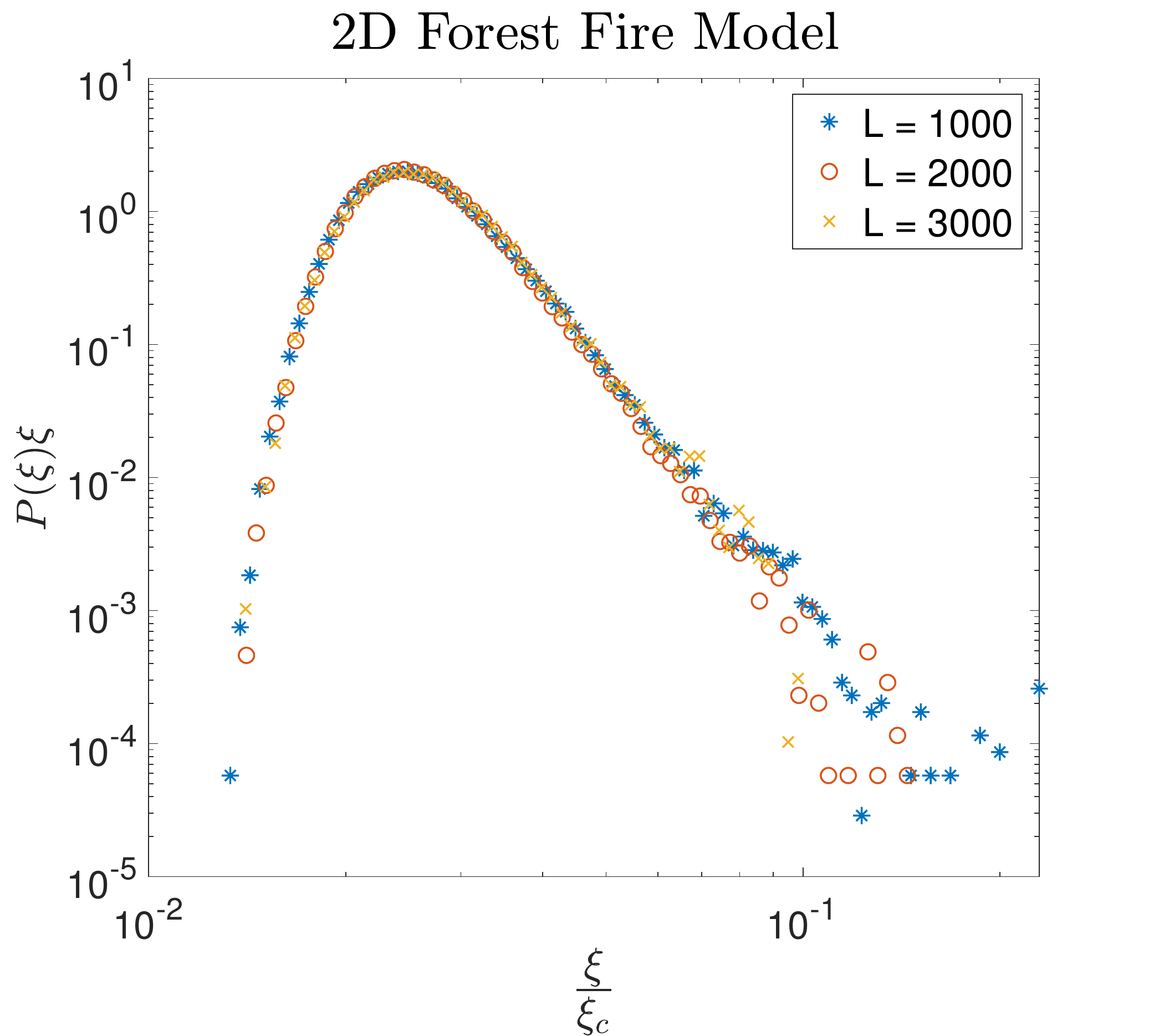}
\caption{Universal function $G(u)$ for the Forest Fire Model with $\frac{\theta}{L^2}=10^{-3}$. The distribution $P(\xi)$ becomes scale invariant in correspondence of $\tau=1$ and $\beta=1.123$.}
\label{FFM}
\end{figure}

\begin{figure}[h]
\centering
\includegraphics[width=9cm,height=9cm]{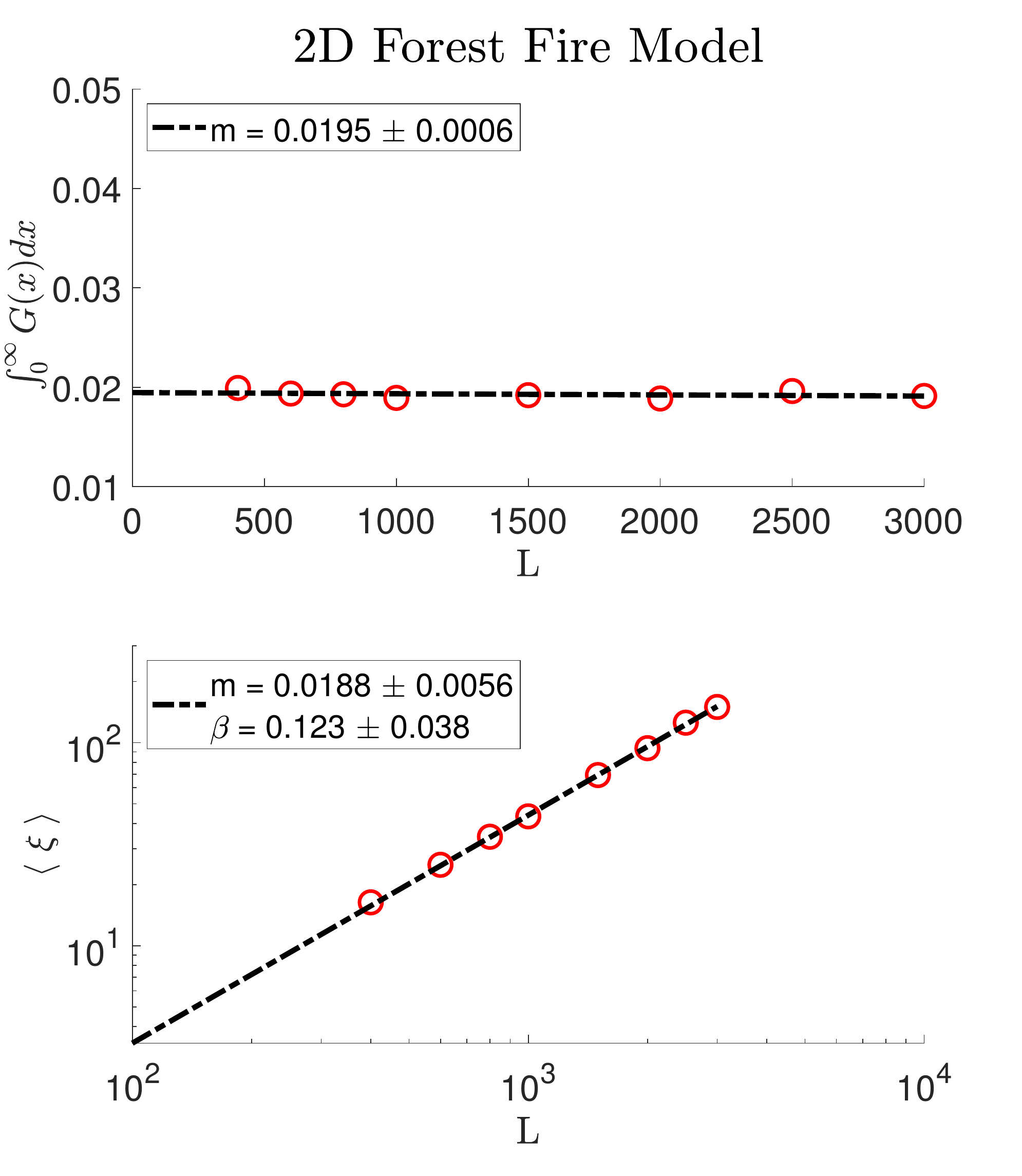}
\caption{{Estimate of the constant of proportionality $m$ between $\langle \xi \rangle$ and $L^\beta$. In the upper panel, $m$ is estimated fitting the integral of the universal function $G(x)$ with a line, while in the lower panel it is fitted directly from $\langle \xi \rangle$. In the FFM, $\langle \xi \rangle$ grows algebraically, which results in a straight line with gradient $\beta$ in the log-log plot.  The error corresponds to confidence bounds of $95\%$.}}
\label{FFM_fit}
\end{figure}

As we discussed in the previous section, it is not clear how to tune the system size $L$ and $\theta$. In previous studies on the correlation length, the standard procedure consisted in keeping the system size $L$ fixed and looking at the behavior of the correlation length as a function of $\theta$ \cite{Henley1993, Honecker1997}. Following this approach, it turns out that it is impossible to perform a data collapse for $P(\xi)$, which agrees with the general lack of scaling observed in the literature so far. The same broken scaling can be observed keeping $\theta$ fixed and changing the value of $L$.
\\ If we consider the correlation length as a surface in the space of parameters $\xi(\theta,L)$, to keep one of the two dimensions fixed corresponds to two different ways of crossing this surface. In particular, increasing the systems size $L$ without a suitable re-scaling of the parameter $\theta$ could lead to a different statistical behavior of the system, although most observables like the average density of trees or the average cluster size seem to be quite robust for a wide range of $\theta$ at a fixed $L$. Even though there are infinite ways of coupling $\theta$ and $L$, it is sensible to choose $\frac{\theta}{L^2}=k$ for a constant $k$ ($k=10^-3$ in our simulations). In this way, for different system sizes, one tries to plant the same fraction of trees, which seems to be reasonable if one wants to assure statistical consistency at different values of $L$. This particular path choice is shown in Fig. \ref{surf}. Surprisingly, coupling the value of $\theta$ and $L$ in this way makes a data collapse for $P(\xi)$ possible (Fig. $\ref{FFM}$), making $P(\xi)$ the first scale-invariant distribution observed in the Forest Fire Model so far. As for the Ising Model and the XY Model, we used $\xi_0=1$ and $\xi_c=\frac{L^\beta}{\sqrt{2}}$ , but this time we found $\tau=1$ and $\beta = 1.123\pm 0.038$ (Fig. \ref{FFM_fit}), which corresponds to $\nu = 0.561 \pm 0.019$ with a $95\%$ confidence bound. This measurement is consistent with the exponents that have been computed for the two-point and the connected correlation lengths in previous studies \cite{Henley1993,Honecker1997}. We conclude this section observing how the broken scaling in the distribution of cluster sizes $P(S)$ found in \cite{Grassberger2002, Gunnar2002} is not affected by the choice of keeping fixed the ratio $\frac{\theta}{L^2}$. This means that the distribution of cluster sizes is not scale-invariant, although the distribution of correlation lengths is scale-free. Therefore, even though the clusters grow in a non-critical and non-scale-free way, there seems to be some global order in terms of the correlations, which is highlighted by the scale invariance of $P(\xi)$. This is a highly non-trivial result and an aspect that surely requires further investigations. 
\\ Finally, we observe that the correlations in the FFM seem to grow at a higher rate than in the Ising model and in the XY model ($\beta=1.123$). This is likely due to the burning mechanism, which introduces long range correlations in the system as a consequence of the simultaneous removal of sites that belong to the burning cluster. 

\section{Conclusions}
The instantaneous correlation length formalism that we have introduced was able to reproduce the well-known results about the critical behavior of the Ising Model and the XY Model, proving that $P(\xi)$ can be used to identify the presence of a phase transition and to estimate the asymptotic behavior of the correlation length. Furthermore, the introduction of $P(\xi)$ allowed us to define a new critical exponent $\tau$, which happens to be equal to $1$ for all the three models discussed in the paper. When applied to the Forest Fire Model, this method allowed to identify a coupling of the two parameters $L$ and $\theta$ for which $P(\xi)$ is scale invariant. The scale invariance of $P(\xi)$ was unexpected as it is the first scale-free distribution observed in the model so far, and this opens once again the debate about the criticality of the Forest Fire Model. In particular, we observe that the FFM shares the same critical exponent $\tau=1$ of the Ising model and the XY model but displays an algebraic growth of the correlation length $\beta=1.12$ which could be the reason behind the broken scaling observed in the distribution of cluster sizes $P(S)$.
\\ From a theoretical perspective, all systems that present a critical exponent $\tau=1$ share a very elegant property, namely the fact that constant of proportionality and the system size dependence are described by the integral of the universal function $G(u)$. In the case of the Ising model and the XY model, we found $\tau=1$ and $\beta=1$. This means that all the details of the two models are contained in the integral of the universal function $G(u)$, which is characteristic of the model under analysis and becomes the only relevant quantity to distinguish between the critical behavior of correlations for the Ising model and the XY model. In the appendix, it is discussed in more detail the relationship between the ensemble correlation length and the instantaneous correlation length, and how it is possible to obtain the classic critical exponent for the correlation length starting from the instantaneous correlation length formalism.
\\ Finally, we observe how the presented method could be easily applied to the study of real-world phenomena, such as brain activity or rain precipitation, as the estimate of $P(\xi)$ only requires to collect different images of the system during its time evolution. The study of $P(\xi)$ in real-systems could be a useful tool to assess the scale-invariance of the systems under examination and to contribute to a more accurate characterization of their critical behavior.
\section{Acknowledgment}
LP gratefully acknowledges an EPSRC-Roth scholarship from the Department of Mathematics at Imperial College London, the High-Performance Computing facilities provided by the Research Computing Service, and Gunnar Pruessner for very helpful conversations.
\section{Author contributions}
 Both authors discussed the results of the numerical simulations and contributed to the final version of the manuscript. L. P. performed the numerical simulations and wrote the paper. 
 \appendix
\section{Appendix: Critical exponent of the correlation function}
\begin{figure}[h]
\centering
\includegraphics[width=9cm,height=9cm]{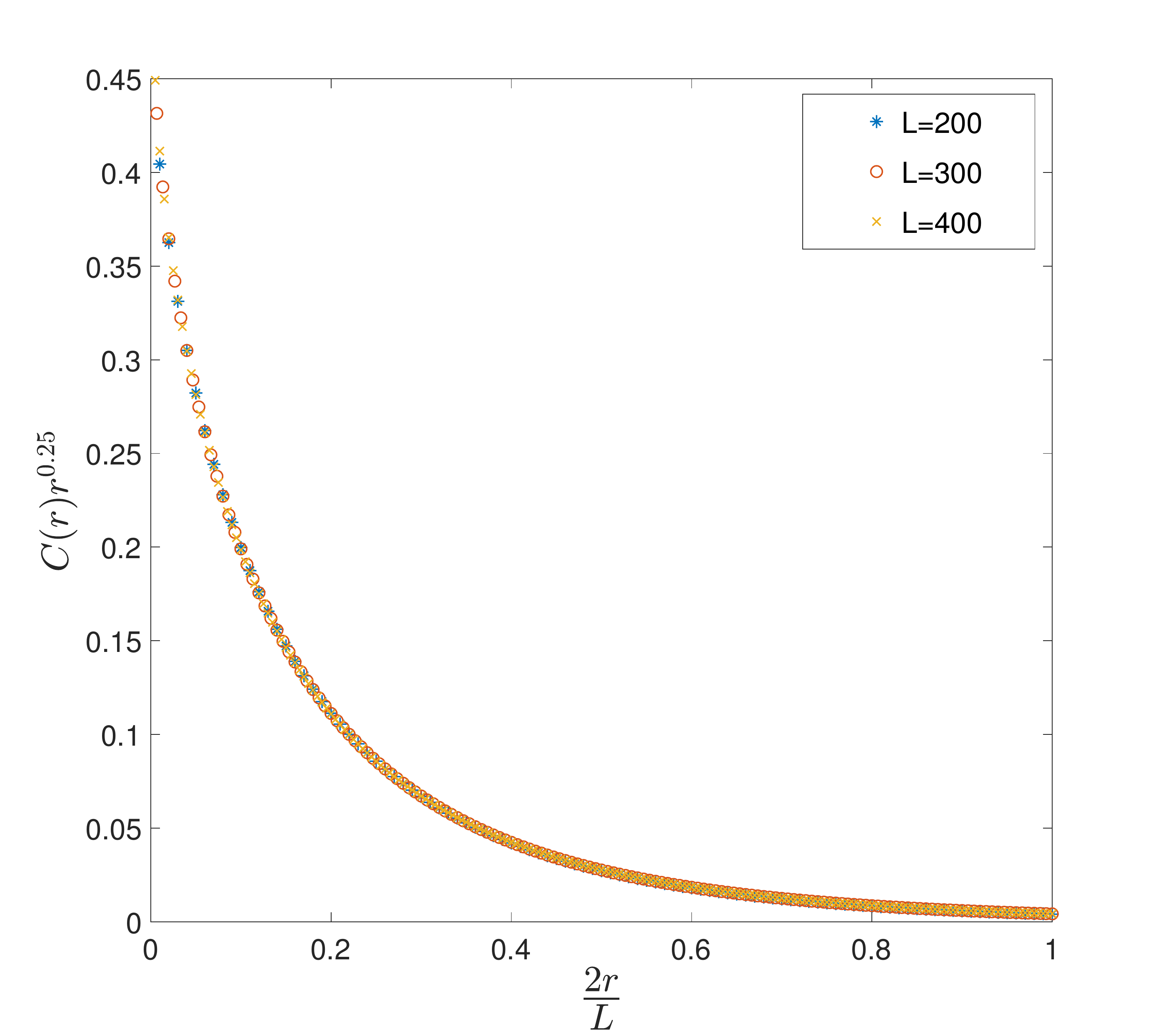}
\caption{It is possible to recover $\eta=0.25$ by reverse engineering the individual correlation functions from the fitted parameters and then performing a data collapse for different values of $L$.}
\label{IsingCorr}
\end{figure}

\begin{figure}[h]
	\centering
	\includegraphics[width=9cm,height=9cm]{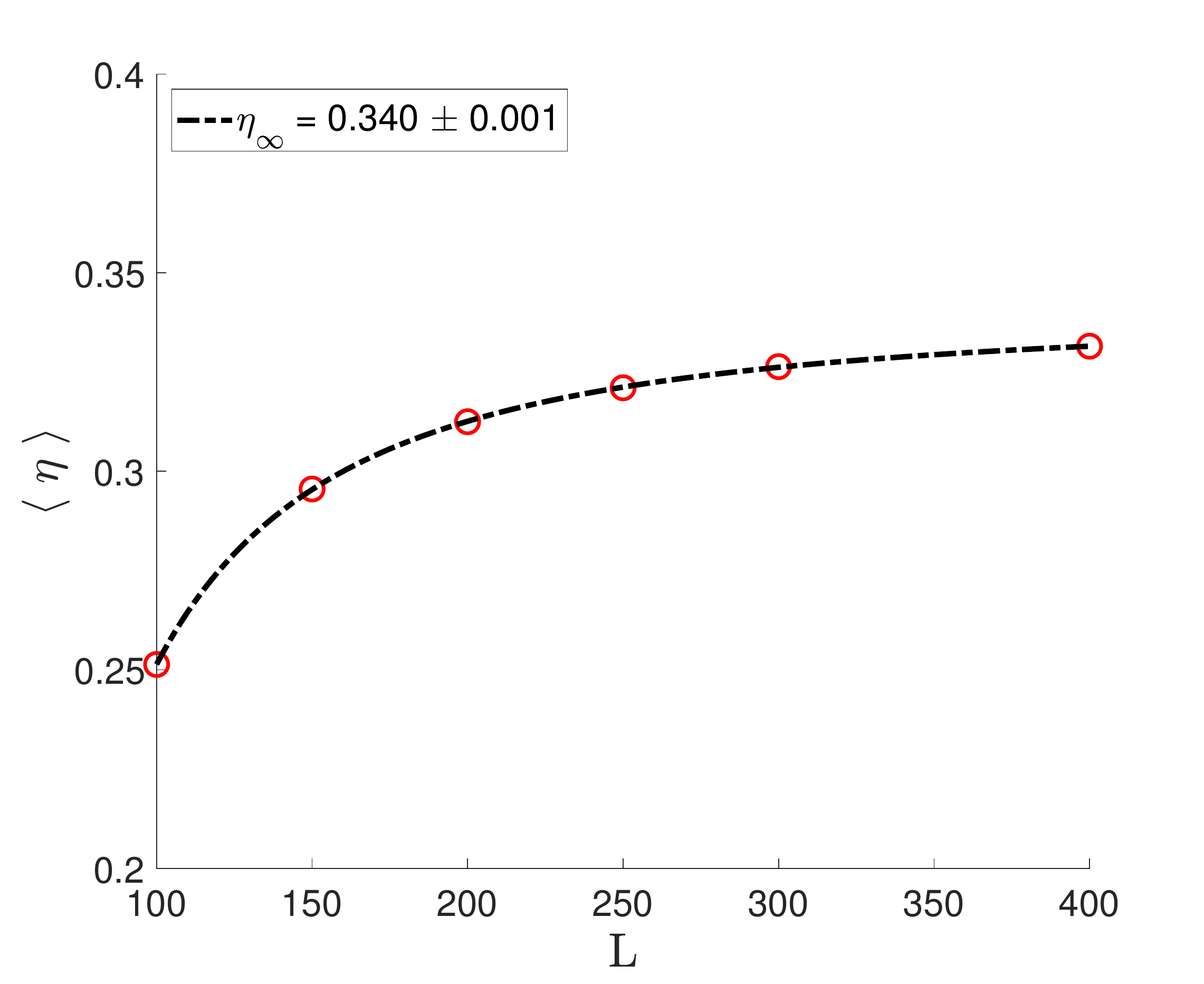}
	\caption{The estimate of $\langle \eta \rangle$ converges towards an asymptotic value of $\langle \eta \rangle = 0.34$ as $L^{-\lambda}$ with $\lambda=1.69 \pm 0.05$. The error corresponds to confidence bounds of $95\%$.}
	\label{aveta}
\end{figure}

As it is well known from classical statistical mechanics, the correlation function of the 2D Ising model is characterized by a critical exponent $\bar{\eta}=0.25$ \cite{chaikin1995}. It is therefore natural to investigate whether it is possible to recover this critical exponent employing the formalism we have hereby introduced. It is worth to stress the fact that although we assume the same functional form for the instantaneous correlation function and the classic one, the instantaneous values of $\xi$ and $\eta$ represent two different mathematical quantities with respect to their traditional counterpart. The crucial point is that we expect the standard correlation length $\bar{\xi}$ and $\xi$ to scale in the same way, even though the two quantities are defined differently. In particular, $\xi$ is a variable that is related to how correlated a single configuration is, and it is not bounded by the system size $L$. Regarding the critical exponent $\eta$, since it is a constant, it is not expected to scale with the system size, and we expect it to converge to a value that could be different from $\bar{\eta}=0.25$ because the two quantities are averaged differently. This is confirmed by our simulations, which show that the distribution of $\eta$ is not scale-invariant and that the mean value of $\eta$ tends to $\langle \eta \rangle =0.34$ as $L$ increases (Fig.\ref{aveta}). However, it is still possible to estimate the ensemble critical exponent $\bar{\eta}=0.25$ and, at the same time, check the accuracy of our method. In order to do so, one can use the parameters estimated via fit for each configuration $i$ and reconstruct the correspondent correlation function $C_i(r)$. If the error that we do in fitting $C_i(r)$ is negligible, we should be able to compute the classical correlation function averaging over all configurations, and hence recover $\bar{\eta}=0.25$. Indeed, plotting $C(r)r^{0.25}$ vs $\frac{2r}{L}$ for different system sizes we can perform a data collapse (Fig.\ref{IsingCorr}), meaning that the fitting error is negligible and that we can safely recover the ensemble critical exponent $\bar{\eta}=0.25$.

\bibliography{bibX}

\end{document}